\documentclass[aps,prb,reprint,amsmath,amssymb,floatfix,superscriptaddress]{revtex4-2}

\makeatletter
\let\frontmatter@footnote@produce\frontmatter@footnote@produce@endnote
\makeatother

\usepackage{float}
\usepackage{lipsum}
\usepackage{titlesec}
\usepackage{natbib}
\usepackage{graphicx}
\usepackage{physics}
\usepackage{bm}  
\usepackage{amsmath} 
\usepackage{braket} 
\usepackage[colorlinks,allcolors=blue,urlcolor=blue]{hyperref}

\usepackage{graphicx}
\usepackage{dcolumn}

\begin{document}

\preprint{APS/123-QED}

\title{\textit{Ab initio} electron dynamics in high electric fields:\\ 
accurate predictions of velocity-field curves
}

\author{Ivan Maliyov}
\author{Jinsoo Park}
\author{Marco Bernardi}%
\email{bmarco@caltech.edu}
\affiliation{%
 Department of Applied Physics and Materials Science, \protect\\ California Institute of Technology, Pasadena, California 91125
}


\begin{abstract}
Electron dynamics in external electric fields governs the behavior of solid-state electronic devices. First-principles calculations enable precise predictions of charge transport in low electric fields.
However, studies of high-field electron dynamics remain elusive due to a lack of accurate and broadly applicable methods. 
Here we develop an efficient approach to solve the real-time
Boltzmann transport equation with both the electric field term and \textit{ab initio} electron-phonon collisions. 
These simulations provide field-dependent electronic distributions in the time domain, allowing us to investigate both transient and steady-state transport in electric fields ranging from low to high ($>10$ kV/cm). 
The broad capabilities of our approach are shown by computing nonequilibrium electron occupations and velocity-field curves in Si, GaAs, and graphene, obtaining results in quantitative agreement with experiment. 
Our approach sheds light on microscopic details of transport in high electric fields, including dominant scattering mechanisms and valley occupation dynamics. Our results demonstrate quantitatively accurate calculations of electron dynamics in low-to-high electric fields, with broad application to power- and micro-electronics, optoelectronics, and sensing.

\end{abstract}



\maketitle

\titlespacing\section{-10pt}{04pt plus 2pt minus 2pt}{2pt plus 2pt minus 2pt}
\titlespacing\subsection{0pt}{12pt plus 4pt minus 1pt}{0pt plus 2pt minus 1pt}

\section{\label{sec:intro}INTRODUCTION}
\vspace{10pt}
The advent of nanoscale transistors and power electronics has made high electric fields widespread in modern devices~\cite{smithe2018high,nathawat2020transient,jaroszynski2008comparative,yang2010above,verzellesi2013efficiency,jaros2019fullerene}. As a result, accurate modeling of high-field electrical transport is broadly relevant for various technologies. 
Electrical transport is often characterized by the velocity-field curve, which describes the mean drift velocity of the charge carriers as a function of applied electric field. 
The drift velocity typically increases linearly at low field, with a slope equal to the carrier mobility, and then saturates at high fields~\cite{meric2008current,yao2000high}.
From a microscopic viewpoint, transport near room temperature is controlled by the interactions between electrons and lattice vibrations (phonons), whereas at lower temperatures or high doping, defects and impurities also play a role. 
Detailed knowledge of the microscopic mechanisms governing the mobility and saturation velocity is important to advance electronic devices and search for improved electronic materials. 
\\
\indent 
Studies of velocity-field curves date back to the early days of semiconductors~\cite{shockley1951hot,gunn1956vi,butcher1965intervalley,johnson1991physical,ferry1975high}. These measurements are now routine, with possible challenges due to sample self-heating~\cite{nathawat2020transient} or spurious substrate effects~\cite{meric2008current}. Theory and computation can aid the interpretation of transport experiments and shed light on the mechanisms limiting the mobility and saturation velocity. The Monte-Carlo (MC) method~\cite{lundstrom1997elementary,joshi2003monte} has been the de facto standard for velocity-field curve calculations since its inception in the 1970s~\cite{littlejohn1977velocity}. 
Semi-empirical MC uses electron interactions that are modeled analytically or fit to experimental data. It is a valuable and versatile tool, but it typically requires a large number of empirical parameters, including carrier effective masses, deformation potentials to describe electron-phonon ($e$-ph) interactions, dielectric properties, and phonon energies~\cite{joshi1994simulations,li2000monte,chauhan2009high,shishir2009velocity}. Overall, MC is not geared toward quantitative predictions, especially in new materials where extensive experimental data is missing.
\\ 
\indent
\textit{Ab initio} methods based on \mbox{density functional theory} (DFT)~\cite{burke2012perspective} have enabled accurate calculations of the electronic structure, phonon dispersions, and $e$-ph interactions~\cite{martin2020electronic,dreizler2012density,murray2007phonon,baroni1987green,baroni2001phonons,gonze1995adiabatic}. Yet, MC studies leveraging these techniques are still uncommon~\cite{ghosh2017ab,akturk2009high,mandal2014strong}, 
and computing velocity-field curves entirely from first principles remains an open challenge.
The Boltzmann transport equation (BTE) provides a convenient framework to study low-field transport and the phonon-limited mobility~\cite{pizzi2014boltzwann, Li2015, Zhou2016, Jhalani2017, Chen2017, Sohier2018, Li2018, Lee_2018, lee2020ab, Jhalani-quad,park2014electron,liu2017first}. These calculations combine electronic and phonon data from DFT with dielectric screening and $e$-ph interactions from density functional perturbation theory (DFPT)~\cite{baroni2001phonons}, providing a seamless workflow to model electrical transport. 
Recent developments enable calculations of transport in magnetic fields~\cite{macheda2018magnetotransport,desai2021magnetotransport} and in the presence of polaron effects~\cite{zhou2019predicting}. 
However, an \textit{ab initio} approach for transport in high electric fields and velocity-field curves is still missing.
\\ 
\indent
\begin{figure*}
\includegraphics[width=1.9\columnwidth]{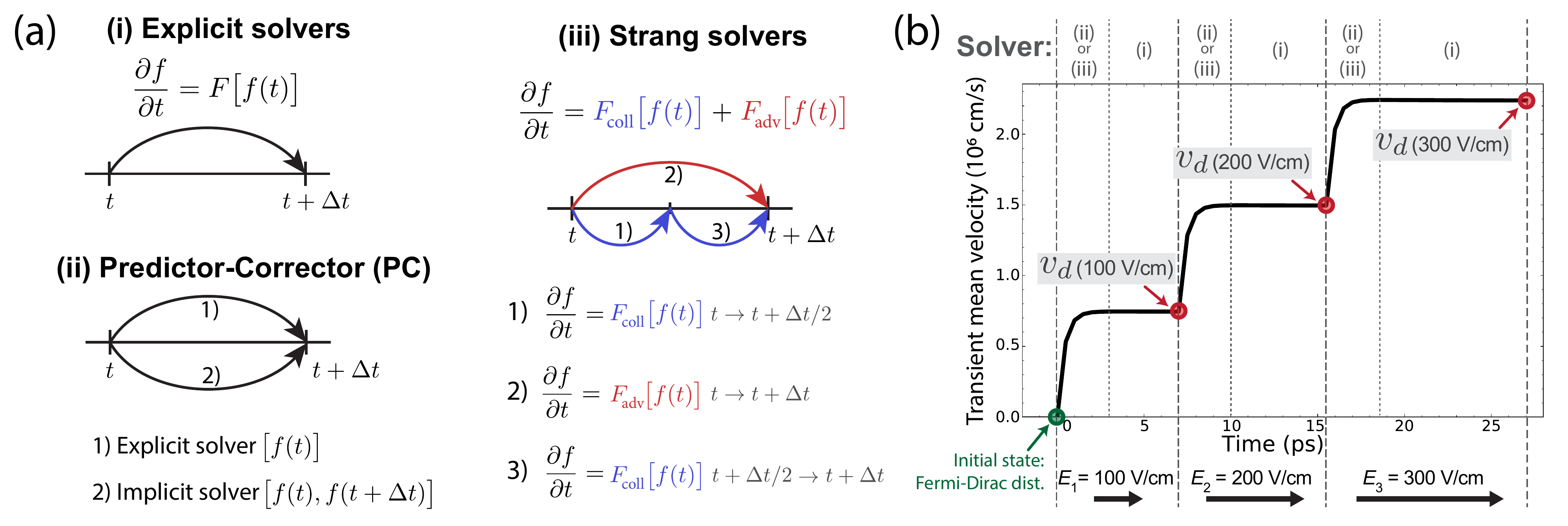}
\caption{
\label{fig:bte_diagram}
(a) Schematic representation of the rt-BTE explicit time-stepping algorithms. (b) Computation of the velocity-field curve, with GaAs results shown as an example. 
} 
\end{figure*}
Here we show a method based on the real-time Boltzmann transport equation (rt-BTE)~\cite{jhalani2017ultrafast,zhou2021perturbo} to accurately predict the velocity-field curve and high-field transport from first-principles. 
The explicit time-stepping of the electron occupations, as achieved in this work, provides access to the real-time electronic response to the field, including both transient and steady-state dynamics.
We develop a range of numerical solvers to time-step the rt-BTE with \textit{ab initio} $e$-ph collisions in the presence of an external electric field. 
From the resulting time-domain electron dynamics, we extract transient and steady-state drift velocities, nonequilibrium electron distributions, and velocity-field curves.  
Our method, applied here to Si, GaAs, and graphene, is shown to provide velocity-field curves in excellent agreement with experiments 
(without using free parameters) from low field up to saturation. Analysis of electron distributions and phonon scattering mechanisms as a function of electric field allows us to access microscopic details of high-field transport. Taken \mbox{together}, this work paves the way for quantitative \textit{ab initio} studies of velocity-field curves and high-field transport phenomena.\\
\vspace{10pt}
\section{\label{sec:method} METHODS}
%
\subsection{\label{subsec:rt-bte} Real-time Boltzmann Transport Equation}
\vspace{0.1cm}

The rt-BTE for a homogeneous material in the presence of an external electric field reads~\cite{mahan2010condensed}
\begin{equation}
\label{eq:rt-bte}    
\frac{\partial f_{n\mathbf{k}}(t)}{\partial t} = \frac{e\mathbf{E}}{\hbar}\cdot \nabla _\mathbf{k}f_{n\mathbf{k}}(t)+\mathcal{I}^{e-\mathrm{ph}}[f_{n\mathbf{k}}(t)],
\end{equation}
where $f_{n\mathbf{k}}(t)$ is the time-dependent electronic occupation of a Bloch state with band index $n$ and crystal momentum $\mathbf{k}$, and $e$ is the electronic charge. The first term on the right-hand side, here called the advection term, is proportional to the external electric field $\mathbf{E}$ and is responsible for electron drift. The second term is the $e$-ph collision integral $\mathcal{I}^{e-\mathrm{ph}}$ accounting for phonon absorption and emission processes, which restore equilibrium and drive the electronic occupations to a steady state~\cite{ziman2001electrons,mahan2010condensed,bernardi2016first}. Its explicit expression is given in Eq.~(2) of Ref.~\cite{zhou2021perturbo}.
\\
\indent
The $e$-ph collision integral depends on the electronic occupations, so the rt-BTE is a set of coupled integro-differential equations, whose numerical solution is challenging due to computational cost~\cite{zhou2021perturbo,tong2020precise}. In the presence of an electric field, numerical instabilities complicate the problem further. 
We develop multiple schemes to solve the rt-BTE in Eq.~(\ref{eq:rt-bte}) 
by explicit time-stepping, using a uniform time grid, $t=n\Delta t$, with a time step $\Delta t$ of order 1~fs. 
We divide the solvers into three groups: Solvers in groups (i) and (ii) treat the right-hand side of Eq.~\eqref{eq:rt-bte} as a single term $F\big[f(t)\big]$, and solvers in group (iii) consider separately the advection and collision terms, denoted as $F_{\mathrm {adv}}\big[f(t)\big]$ and $F_{\mathrm{coll}}\big[f(t)\big]$ respectively.
\\
\indent
Group (i) contains two explicit solvers, forward Euler (FE) and 4th-order Runge-Kutta (RK4), which calculate the occupations at the next time step, $f_{n\mathbf{k}}(t+\Delta t)$, using only the values $f_{n\mathbf{k}}(t)$ at the current time $t$~\cite{abramowitz1964handbook}. 
Group (ii) includes predictor-corrector (PC) solvers~\cite{abramowitz1964handbook,diethelm2002predictor} that first predict $f_{n\mathbf{k}}(t+\Delta t)$ with an explicit method and then correct it with an implicit step using both $f_{n\mathbf{k}}(t)$ and the predicted $f_{n\mathbf{k}}(t+\Delta t)$. 
We use the FE or RK4 solvers for the predictor step and backward Euler (BE) or Crank-Nicolson (CN)~\cite{thomas2013numerical} for the corrector step.
Finally, group (iii) solvers carry out the advection and collision steps separately. In this group, we find optimal results with the Strang splitting technique~\cite{strang1968construction}, which advances the rt-BTE to the next time step using a three-step sequence $-$ half-step advection followed by full-step collision and half-step advection (ACA solver), or the same step sequence but with collision-advection-collision order (CAC solver). Each step can be performed with any of the solvers in groups (i) and (ii).  
\\
\indent
These three groups of solvers are shown schematically in Fig.~\ref{fig:bte_diagram}(a). Detailed equations for all solvers are given in the Supplemental Material~\cite{supplemental}.
We have tested a total of 78 nontrivial solver combinations, including a range of PC and Strang splitting schemes. 
On the basis of accuracy and computational cost, optimal solvers are FE for group (i), FE predictor plus BE corrector for group (ii), and for group (iii), CAC with the best PC in group (ii) for the advection term and FE for the collision term.  
Solvers in groups (ii) and (iii) are more accurate and stable, but require more stages per time step and thus are more computationally expensive. 

%
\subsection{\label{subsec:vel_field_procedure} Velocity-field curve calculations}

The drift velocity quantifies the steady-state motion of the charge carriers in the direction of the applied electric field. Using the electronic occupations $f_{n\mathbf{k}}(t)$ from the rt-BTE, we compute a transient mean velocity parallel to the field at each time step:
\begin{equation}
\label{eq:drift_vel}
\langle v_{\|}(t) \rangle = \frac{1}{n_\mathrm{c} \mathit{\Omega} \mathcal{N}_\mathbf{k}} \sum_{n\mathbf{k}} f_{n\mathbf{k}}(t) \, (\mathbf{v}_{n\mathbf{k}}\cdot \hat{\mathbf{E}}),
\end{equation}
where $\mathbf{v}_{n\mathbf{k}}$ is the band velocity, $\mathbf{v}_{n\mathbf{k}}\cdot \hat{\mathbf{E}}$ its projection along the field direction $\hat{\mathbf{E}}$, $n_\mathrm{c}$ is the carrier concentration, $\mathit{\Omega}$ is the volume of the unit cell, and $\mathcal{N}_\mathbf{k}$ is the number of $\mathbf{k}$-points used in the summation.
To obtain the drift velocity $v_\mathrm{d}$, we compute $\langle v_{\|}(t) \rangle$ at steady state, which is achieved after a long enough simulation time:
\begin{equation}
\label{eq:drift_vel_steady}
v_\mathrm{d} = \langle v_{\|}(t\rightarrow \infty) \rangle.
\end{equation}

Mapping the entire velocity-field curve requires separate calculations of the drift velocity for each value of the electric field, in each case selecting an appropriate starting point for the electronic occupations and running the simulation long enough to reach a steady state. 
We have devised a streamlined procedure to more rapidly reach steady-state distributions and calculate the entire velocity-field curve in a single simulation [see Fig.~\ref{fig:bte_diagram}(b)]. 
Using initial occupations that follow a Fermi-Dirac distribution in thermal equilibrium, we apply a small electric field, $E_1\!=\!100$~V/cm in Fig.~\ref{fig:bte_diagram}(b), to drive the system out of equilibrium, and then time-step the rt-BTE to a steady state and compute the drift velocity. Next, we increase the electric field to a new value, $E_2=200$~V/cm in Fig.~\ref{fig:bte_diagram}(b), and time-step the rt-BTE using as a starting point the steady-state occupations from the previous field value $E_1$. This process of increasing the electric field, time-stepping to a new steady state, and recording the new value of the drift velocity is repeated until reaching the maximum electric field of interest~\cite{supplemental}.
\\
\indent
Each time the value of the electric field is increased, the mean velocity $\langle v_{\|}(t) \rangle$ changes abruptly [Fig.~\ref{fig:bte_diagram}~(b)]. During these transients with rapid changes in the electron occupations, the rt-BTE solver can develop numerical instabilities, here tackled effectively with the solvers in groups (ii) or (iii).
This rapid transient is followed by a long time window with slow velocity changes until a steady state. In this regime, where the rt-BTE is stable and well-behaved, we employ the simple solvers in group (i), such as FE, to more efficiently time-step the rt-BTE to a steady state. 
%
\subsection{\label{subsec:pert} Computational details}
\vspace{5pt}
We carry out plane-wave DFT calculations with the {\sc{Quantum Espresso}} code~\cite{QE2017} to compute the ground state and band structure of Si, GaAs, and graphene, using the local density approximation (LDA)~\cite{perdew1992accurate}, norm-conserving pseudopotentials~\cite{troullier1991efficient}, and relaxed lattice parameters. 
For GaAs, we refine the band structure using $G_0W_0$ calculations with the \textsc{YAMBO} code~\cite{sangalliManybody2019}. 
The phonon dispersions and $e$-ph perturbation potentials are computed with DFPT~\cite{QE2017}. 
The $e$-ph matrix elements are computed on coarse $\mathbf{k}$- and $\mathbf{q}$-point grids (where $\mathbf{q}$ is the phonon wave-vector) with our {\sc{Perturbo}} open-source package~\cite{zhou2021perturbo}, and then interpolated to finer grids using Wannier functions generated with {\sc{Wannier90}}~\cite{Mostofi_2014}.
\\
\indent 
All the rt-BTE solvers discussed above have been implemented in {\sc{Perturbo}}. The $\mathbf{k}$-space gradient in the advection term of the rt-BTE [see Eq.~\eqref{eq:rt-bte}] is computed using a finite difference method~\cite{mostofi2008wannier90}. We time-step the rt-BTE with a 1$-$20~fs time step. After applying a new electric field value, we use the CAC solver for the first 3~ps to improve the stability, and then employ the FE solver to reach a steady state, typically within 10$-$200~ps.
For the rt-BTE dynamics and drift velocity, we use dense, uniform and equal $\mathbf{k}$- and $\mathbf{q}$-point grids together with tetrahedron integration~\cite{blochl1994improved}. The momentum-averaged energy-dependent electronic populations are obtained as $\bar{f}(E,t)=\sum_{\mathrm{n}\mathbf{k}} f_{n\mathbf{k}}(t) \delta(\epsilon_{\mathrm{n}\mathbf{k}} - E)$. Additional numerical details, such as cutoffs and grids, are provided below~\cite{methods-extra}.

\begin{figure}
\centering
\includegraphics[width=0.87\columnwidth]{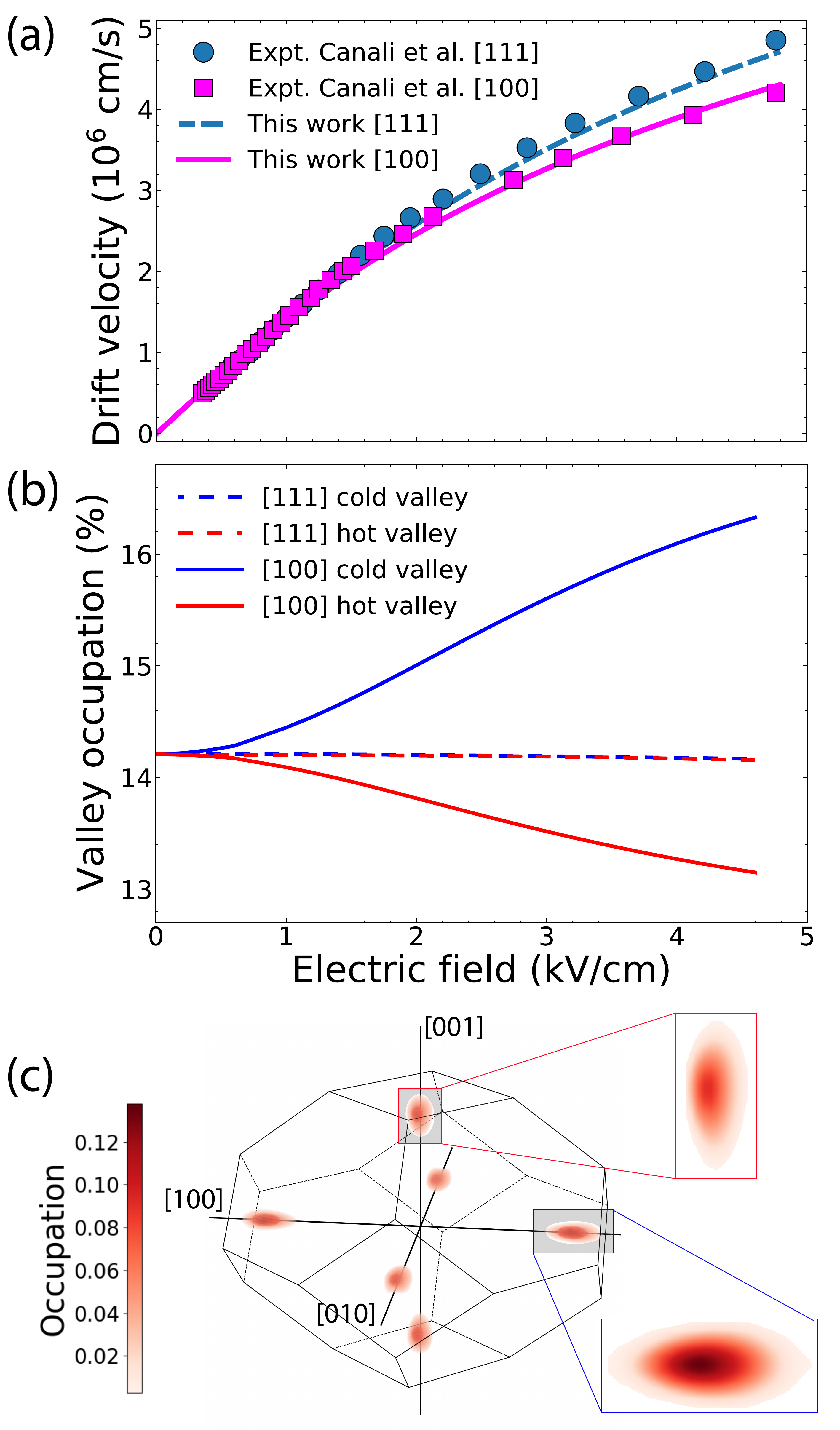}
\caption{Computed results in silicon for an external electric field oriented along the [100] and [111] directions. (a) Velocity-field curves, compared with experimental data from Ref.~\cite{canali1971drift}. (b) Occupation of hot (red) and cold (blue) conduction band valleys as a function of electric field. (c) Electron occupations for a 4600 V/cm electric field along [100], with insets zooming into the hot (top) and cold (bottom) valleys. 
}
\label{fig:vel_field_si}
\end{figure}
\begin{figure*}[t]
\centering
\includegraphics[width=2.05\columnwidth]{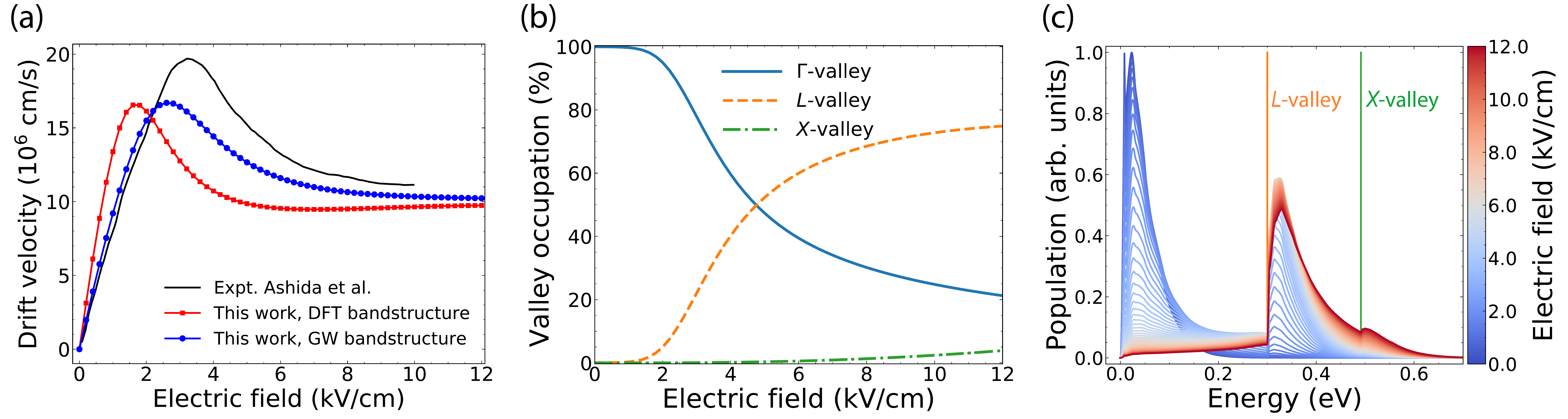}
\caption{(a) Velocity-field curve in GaAs at 300~K computed with DFT and GW band structures and compared with experimental data~\cite{ashida1974energy}. (b) Electronic valley occupations in GaAs as a function of applied electric field. The Brillouin zone regions defining the valleys are obtained from a 0.75 eV energy cutoff above the conduction band minimum. (c) Steady-state electron occupations as a function of conduction band energy and applied electric field. The $L$- and $X$-valley minima  are shown with vertical lines.} 
\label{fig:vel_field_gaas}
\end{figure*}

\vspace{0.2cm}
\section{\label{sec:results} RESULTS}
\vspace{5pt}
%
We present results for Si and GaAs as examples, respectively, of a nonpolar and polar semiconductor, and for graphene as a metal. At low field, the accuracy of the linearized BTE with \textit{ab initio} $e$-ph collisions in these systems has been established in previous work~\cite{park2014electron, Li2015, Zhou2016}. However, much less is known about calculations at high electric fields, where the carrier distributions are driven more extensively out of equilibrium, so the accuracy of the linearized BTE is not guaranteed. 
Note also that our approach of time-stepping the rt-BTE to a steady state, as opposed to solving the linearized BTE~\cite{Li2015,zhou2021perturbo},  
has not been tested even at low fields.
Our rt-BTE method is not limited to small changes in occupations $-$ the main assumption of the linearized BTE $-$ so it allows us to investigate electron dynamics in electric fields ranging from low to high ($>10$ kV/cm).
\\
\indent
\textit{Silicon.} The velocity-field curves in Si calculated from the rt-BTE steady-state occupations are shown in Fig.~\ref{fig:vel_field_si}~(a) for electric fields applied in the [111] and [100] directions. For both directions, we find an excellent agreement with experiment. At low-field, the mobility obtained with the rt-BTE is within 3\% of the experimental value. At higher electric fields, the computed drift velocity accurately follows the experimental trend, providing a drift velocity greater in the [111] than in the [100] direction. This result is a well-known consequence of the Si band structure~\cite{kovi2013charge,canali1971drift}, where the six conduction band valleys possess a greater effective mass $-$ and thus a lower mobility $-$ in the equivalent [100] longitudinal directions than in the transverse directions~\cite{Yu-Cardona}.
\\
\indent
When the electric field is applied in the [111] direction, all six valleys are equivalent and equally excited due to the cubic symmetry. However, for an electric field oriented along [100], two valleys are oriented in the longitudinal direction parallel to the electric field, and four valleys are oriented in the orthogonal (transverse) directions. 
Electrons in these four valleys, due to the smaller transverse effective mass, are excited more extensively, leading to four so-called ``hot" valleys in the transverse, and two ``cold" valleys in the longitudinal directions. The resulting hot-valley depletion increasing with electric field is clearly seen in our numerical results in Fig.~\ref{fig:vel_field_si}~(b). 
For a high electric field of $\sim$4.5~kV/cm, visualization of the valley occupations, shown in Fig.~\ref{fig:vel_field_si}~(c), clearly shows the hot-valley depletion for an electric field applied in the [100] direction. 
%

\textit{Gallium arsenide.} 
The velocity-field curve in GaAs has been studied extensively due to its unusual characteristics. While the drift velocity in GaAs increases linearly at low field, at higher field values it exhibits a peak followed by a region of velocity decrease, which corresponds to a negative differential resistance. 
This trend is often referred to as the Gunn effect~\cite{kroemer1964theory,shaw1980gunn} and is widely used in electronics for microwave generation and sensors~\cite{qi2006spin,xu2008gunn,litovchenko2005gunn,khalid2014terahertz}. 
In the Gunn effect in GaAs, electrons in the $\Gamma$-valley are scattered to the higher-energy, lower-mobility $L$- and $X$-valleys for increasing values of the electric field, resulting in lower drift velocities as more electrons are transferred to the higher-energy valleys. 
\\
\indent
Figure~\ref{fig:vel_field_gaas}(a) shows our velocity-field curve in GaAs computed with the rt-BTE using both DFT and GW band structures. We find that the drift velocity is sensitive to the electronic band structure, which regulates the band velocities and $e$-ph scattering processes. 
The GW calculation significantly improves the band structure, giving an effective mass ($0.069~m_e$ in GW versus $0.049~m_e$ in DFT) and $L$-valley position (0.3 eV above the conduction band minimum) in excellent agreement with experiments~\cite{blakemore1982semiconducting}. 
Compared to DFT, the GW calculation provides a better agreement with experiment, improving the low-field mobility by 70\% and the velocity peak position by 20\% due to the more accurate band structure.
\\
\indent
These improvements allow us to predict the velocity-field curve in GaAs with a high accuracy, as is shown in Fig.~\ref{fig:vel_field_gaas}(a) by comparing our computed GW velocity-field curve with experiments. 
The low-field mobility obtained from the rt-BTE using the GW band structure ($9920\;\mathrm{cm^2/Vs}$) overestimates the experimental values ($6350-9000\;\mathrm{cm^2/Vs}$)~\cite{rode1970electron,blakemore1982semiconducting} due to higher-order $e$-ph scattering processes not included in our rt-BTE~\cite{lee2020ab}.
The computed peak velocity ($16.7$~$\mathrm{cm/s}$) is comparable in our DFT and GW calculations, and is in a very good agreement with the experimental value of $19.6$~cm/s. The high-field saturation velocity also agrees with experiment.
\\
\indent
Our rt-BTE dynamics can capture the progressive occupation of the $L$- and $X$-valleys at increasing electric fields [Fig.~\ref{fig:vel_field_gaas}(b)]. The $L$-valley becomes occupied at fields greater than $2$~kV/cm, corresponding to the onset of the drift velocity peak~\cite{blakemore1982semiconducting}. Due to its higher energy, the $X$-valley gets minimally occupied even at $>\!10$~kV/cm field values. 
The momentum-averaged electron occupations as a function of energy and electric field, shown in Fig.~\ref{fig:vel_field_gaas}(c), reveal in detail this valley occupation dynamics. These results show clearly that our rt-BTE simulations can quantitatively describe high-field transport in \mbox{semiconductors.} 
\begin{figure}
\centering
\includegraphics[width=0.86\columnwidth]{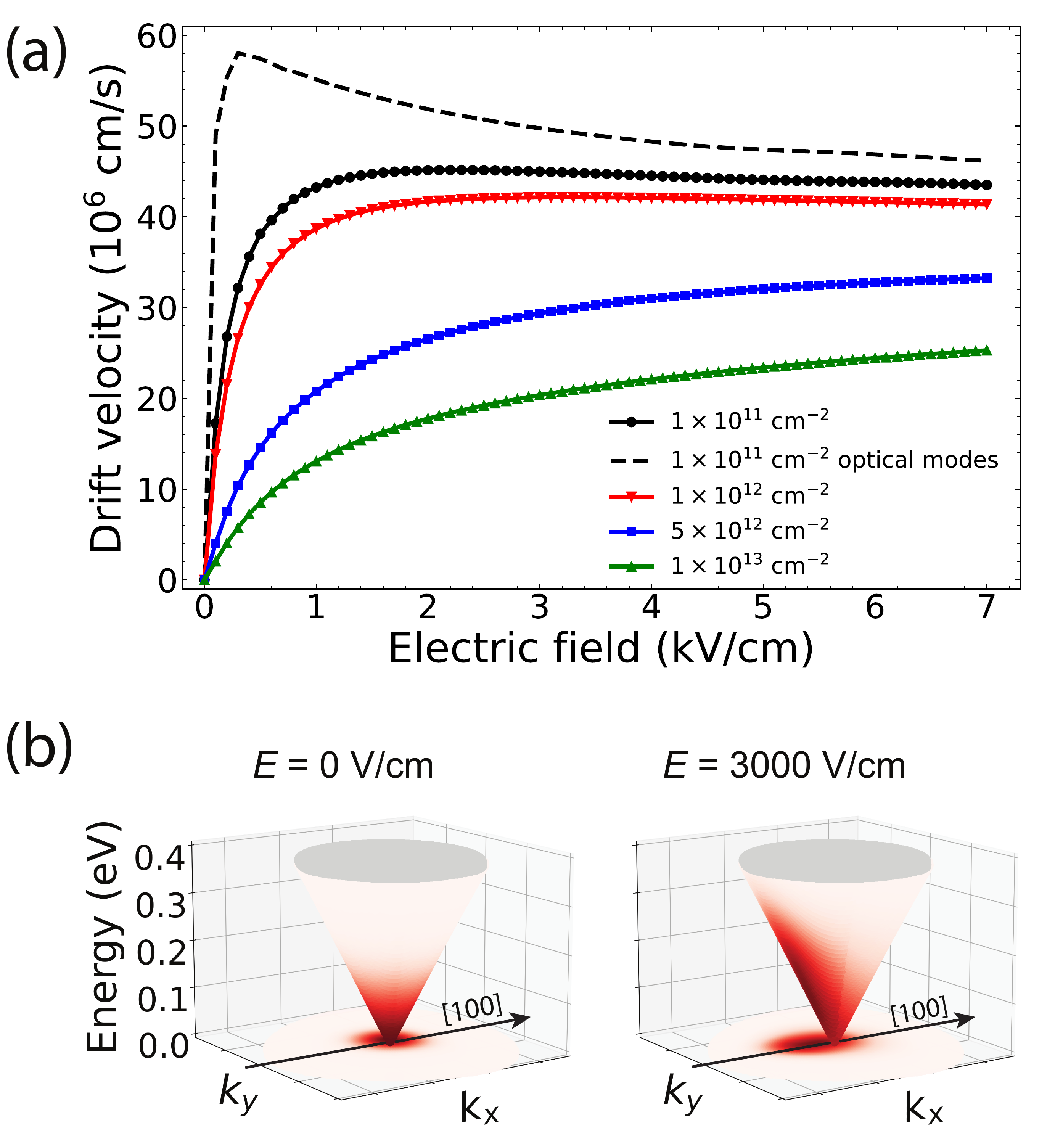}
\caption{(a) Velocity-field curves in graphene for different carrier concentrations. The drift velocity calculated with the optical modes only (for $n=10^{11}\;\mathrm{cm}^{-2}$) is shown with a black dashed line. (b) Electron occupations for a carrier concentration of $n=10^{12}\;\mathrm{cm}^{-2}$ and external electric field of 0 (left) and 3000 V/cm (right).
}
\label{fig:vel_field_graphene}
\end{figure}
%
%

\textit{Graphene.} 
We present results for graphene as an example of a two-dimensional semimetal with exceptionally high mobility~\cite{novoselov2004electric}. Different from GaAs and Si, electrical transport in graphene depends strongly on carrier concentration.
We compute the velocity-field curves for electron concentrations in the $10^{11}$ $-$ $10^{13}\;\mathrm{cm}^{-2}$ range. Our calculations, shown in Fig.~\ref{fig:vel_field_graphene}(a), predict electron mobility values between $0.2 \cdot 10^5$ $-$ $1.5\cdot10^6$~cm$^2$/Vs and saturation velocities between $25\cdot10^{6}$ $-$ $43.5\cdot10^{6}$~cm/s in that range of electron concentrations. 
\\
\indent
Comparing these results with experiments is nontrivial. Our calculations are carried out on ideally pure and isolated graphene in a regime where transport is phonon-limited. However, measurements of the drift velocity in suspended graphene are difficult $-$ experimental results give a wide range of saturation velocities, with significant sample-to-sample variation often attributed to disorder and impurities~\cite{dorgan2013high}. 
The highest measured saturation velocity, $35\cdot10^{6}\;\mathrm{cm/s}$ for a carrier concentration of $4\cdot10^{12}\;\mathrm{cm}^{-2}$, presumably corresponds to a very \lq\lq clean\rq\rq~graphene sample and is in excellent agreement with our predicted value of $30\cdot10^{6}\;\mathrm{cm/s}$ for that carrier concentration. Measurements for substrate-supported graphene are more reliable, but scattering with substrate phonons, not considered here, is known to be important. 
\\
\indent
In graphene, the drift velocity saturates at relatively low electric fields~\cite{meric2008current,dorgan2010mobility,dorgan2013high}, limiting devices applications. 
To study the role of $e$-ph scattering due to optical phonons, we recalculated the velocity-field curve at low carrier concentration including only scattering with optical phonons. 
This result, given in Fig.~\ref{fig:vel_field_graphene}(a), shows that at low electric fields scattering with optical phonons is negligible, consistent with the conventional wisdom that the mobility in graphene is limited by acoustic phonons~\cite{kaasbjerg2012unraveling,hwang2008acoustic,fang2011high}. However, at higher field values, the curve computed with optical phonon scattering only agrees to within less than 10\% with the calculation including all phonon modes. This result demonstrates unambiguously that the saturation velocity is limited by scattering between electrons and optical phonons.
\\
\indent
Comparison of the electron occupations at zero and high field values [Fig.~\ref{fig:vel_field_graphene}(b)] demonstrates that at high field (3~kV/cm, well in the saturation regime), the electrons still occupy mainly the Dirac cones, but their distribution becomes elongated in reciprocal space along the direction of the applied electric field. We conclude that the high-energy tails of this distribution are responsible for the dominant optical phonon emission governing the saturation velocity~\footnote{Finally, note that our drift velocities might be slightly underestimated due to the lack of electron-electron interactions~\cite{fang2011high}, which are expected to be important at very high fields.
Also, it was shown that the electron-two-phonon scattering by the out-of-plane flexural phonon modes has an important contribution to the transport in suspended graphene in the absence of tension \cite{castro2010limits,morozov2008giant,mariani2010temp}. However, the two-phonon scattering processes are not considered here. The inclusion of these effects will be addressed in future work.
}.\\

\section{CONCLUSION}
%
We developed first-principles calculations of electron dynamics in an applied electric field by explicitly time-stepping the rt-BTE with a combination of numerical solvers. Our results establish the rt-BTE framework as an accurate and versatile approach to study high-field transport and seamlessly compute velocity-field curves from first principles.
Including the electric field term in the recently developed rt-BTEs for coupled electron and phonon dynamics~\cite{caruso2021nonequilibrium,tong2020precise} is a goal for future work.
We plan to make the electric field rt-BTE approach available in our open-source code {\sc{Perturbo}}~\cite{zhou2021perturbo} to equip the community with reliable calculations of transport in high electric fields. Taken together, our work expands the first-principles toolbox for studying electron dynamics in real materials and provides an alternative to Monte Carlo calculations for quantitative studies of velocity-field curves and high-field transport. 
\\
\indent

\begin{acknowledgments}
I.M. acknowledges the support by the Liquid Sunlight Alliance, which is supported by the U.S. Department of Energy, Office of Science, Office of Basic Energy Sciences, under Award Number DE-SC0021266. J.P. acknowledges support by the Korea Foundation for Advanced Studies. This research used resources of the National Energy Research Scientific Computing Center (NERSC), a U.S. Department of Energy Office of Science User Facility located at Lawrence Berkeley National Laboratory, operated under Contract No. DE-AC02-05CH11231. 
\end{acknowledgments}

\bibliographystyle{apsrev4-2}

\end{document}